\documentclass[lettersize,journal]{IEEEtran}
\usepackage{amsmath,amsfonts}
\usepackage{amssymb}        % For additional math symbols        % for [H]
\usepackage{algorithm}        % floating environment
\usepackage{algpseudocode}    % For the algorithmic commands
\usepackage{array}
\usepackage{textcomp}
\usepackage{stfloats}
\usepackage{url}
\usepackage{verbatim}
\usepackage{graphicx}
\usepackage{subfigure}
\usepackage{xcolor}
\usepackage{float}
\def\BibTeX{{\rm B\kern-.05em{\sc i\kern-.025em b}\kern-.08em
    T\kern-.1667em\lower.7ex\hbox{E}\kern-.125emX}}
\usepackage{balance}

\floatname{algorithm}{\textbf{Algorithm}}

\begin{document}
% \title{High scale downscaling}
% \title{Efficient Image Downscaling for High Scaling-Factor}
% \title{Revisiting Co-occurrence Learning for Image Downscaling By High Scaling Factor}
% \title{Large Scale  Image Downscaling via Co-occurrence Learning}
%\title{Large Scale  Image Downscaling Using Pixel Bigram Map}
%\title{Revisited Co-occurrence Learning for Image Downscaling With High Scaling Factor}
%\title{An Image Downscaling Method for High Scaling Factor}
% 
%\title{High-Quality and Large-Scale Image Downscaling: \\ Application to Efficient Image Communication}
% \title{Efficient Image Downscaling At High Scale Factor}
\title{High-Quality and Large-Scale Image Downscaling \\ for Modern Display Devices}

% LSID: large-scale image downscaling

\author{Suvrojit Mitra, G B Kevin Arjun, and Sanjay Ghosh, \IEEEmembership{Senior Member, IEEE}
\thanks{This research is supported by the Faculty Start-up Research Grant (FSRG), IIT Kharagpur, awarded to Dr. Sanjay Ghosh.}
\thanks{Suvrojit Mitra, G B Kevin Arjun, and Sanjay Ghosh are with Department of Electrical Engineering, Indian Institute of Technology Kharagpur, WB 721302, India (e-mail: \texttt{sanjay.ghosh@ee.iitkgp.ac.in}).}
}

% \markboth{IEEE Transactions on Consumer Electronics,~Vol.~XX, No.~X, XXX~202X}%
% {How to Use the IEEEtran \LaTeX \ Templates}

% \markboth{IEEE Transactions on Information Forensics and Security,~Vol.~XX, No.~X, XXX~202X}
% {How to Use the IEEEtran \LaTeX \ Templates}

\markboth{Under-Review}%
{How to Use the IEEEtran \LaTeX \ Templates}

\maketitle

\begin{abstract}
In modern display technology and visualization tools, downscaling images is one of the most important activities. This procedure aims to maintain both visual authenticity and structural integrity while reducing an image's dimensions at a large scale to fit the dimension of the display devices. In this study, we proposed a new technique for downscaling images that uses co-occurrence learning to maintain structural and perceptual information while reducing resolution. The technique uses the input image to create a data-driven co-occurrence profile that captures the frequency of intensity correlations in nearby neighborhoods. A refined filtering process is guided by this profile, which acts as a content-adaptive range kernel. The contribution of each input pixel is based on how closely it resembles pair-wise intensity values with it's neighbors. We validate our proposed technique on four datasets: DIV2K, BSD100, Urban100, and RealSR to show its effective downscaling capacity. Our technique could obtain up to $39.22$ dB PSNR on the DIV2K dataset and PIQE up to $26.35$ on the same dataset when downscaling by $8\times$ and $16\times$, respectively. Numerous experimental findings attest to the ability of the suggested picture downscaling method to outperform more contemporary approaches in terms of both visual quality and performance measures. 
Unlike most existing methods, which did not focus on the large-scale image resizing scenario, we achieve high-quality downscaled images without texture loss or edge blurring. 
Our method, LSID (large scale image downscaling), successfully preserves high-frequency structures like edges, textures, and repeating patterns by focusing on statistically consistent pixels while reducing aliasing and blurring artifacts that are typical of traditional downscaling techniques.
\end{abstract}


\section{Introduction}
\label{sec:intro}
\IEEEPARstart{T}{he} utilization of high-resolution images has been augmented as a result of advancements in display technology. These images are frequently downscaled to optimize their appearance, storage, and transmission bandwidth. With the challenge of maintaining structural and textural details, image downscaling is essential for applications like compression \cite{liu2019downscaling}, remote sensing \cite{sdraka2022deep}, video processing \cite{hwang1997high,tan2004fast,kim2018exploiting}, rendering \cite{low2017multi}, and streaming \cite{liu2014content}.An effective downscaling technique helps preserve the important edges that were present in the original image (before the resizing operation), according to a study on the impact of clever edge detection on downscaled photos \cite{kim2020study, zha2023conditional}. 

Conventional techniques such as bilinear and bicubic interpolation \cite{keys1981cubic} may result in aliasing and blurring. Despite sharpening of the images, Lanczos filters \cite{lanczos1964interpolation} can produce ringing artifacts, particularly at lower resolutions. Although co-occurrence learning \cite{ghosh2023image} has better texture information captured by co-occurrence frequency, regularization \cite{liu2018l0} has shown gains in retaining structural details while downscaling.

Despite using edge information to improve interpolation, edge-aware algorithms \cite{canny1986edge, marr1980theory} still struggle with complex textures. 
Although directed filtering \cite{he2013guided} helps focus on important areas, it struggles to balance texture detail and sharpness. 
Although it improves structure preservation, edge-guided interpolation \cite{xu2009edge} has difficulty maintaining consistency of the texture. Although they may not generalize well to a variety of scaling factors, adaptive downscaling techniques \cite{kopf2013content, weber2016rapid} try to strike a balance between efficiency and preservation of detail.

CNNs and auto-encoders are examples of deep learning techniques that can learn mappings from high to low resolution images \cite{dong2016image, vincent2008denoising}, but they need a lot of resources and large datasets. Performance is improved by methods such as residual learning-based denoising CNNs \cite{zhang2018beyond}, but texture preservation and non-integer scaling remain challenges. Although SRGAN \cite{ledig2017photo} and other techniques based on GAN provide aesthetically pleasing results, they can add artifacts, particularly at lower resolutions. Although texture and edge preservation are still difficult, multiscale networks, such as pixel-level interpolation \cite{yang2007downscaling}, improve quality across many scales.

Recent developments include the introduction of an autoencoder framework by Kim et al. that optimizes performance for restoration tasks by concurrently learning downscaling and upscaling networks \cite{kim2018task}. To improve real-world applicability, Xing et al. created a flexible network that can handle any scale factors in a reversible way \cite{xing2023scale}. The authors of \cite{zhang2023adaptive} provided a workable solution for real-time image processing by putting forth a technique for concurrently optimizing rate, accuracy, and latency. IDA-RD, a new metric for quantitatively assessing downscaling techniques, was most recently presented by Liang et al. \cite{liang2024deep}.
Using the notion that downscaling and super-resolution (SR) can be thought of as the encoding and decoding processes in the rate-distortion model, the authors in \cite{liang2024deep} highlighted the significance of an effective image downscaling technique while putting forth a novel metric to quantitatively assess image downscaling algorithms.

%\textcolor{red}{[This intro is just copied from SAID paper.]}

\textbf{Contributions:}
In this paper, we present a downscaling approach that integrates kernel-based filtering with co-occurrence learning, allowing for the tunability of co-occurrence similarity to preserve fine details. Initially, we produce an intermediate image from the input image using a uniform averaging kernel, specifically box filtering, which serves as an approximation of the downscaled image. Subsequently, we assess co-occurrence profile learning utilizing the input image to analyze the frequency of intensity pair occurrences inside a spatial area. Ultimately, we implement a co-occurrence guided kernel filtering that ensures pixels in the input vicinity with robust co-occurrence links to the guide pixel exert a greater influence on the output, thus maintaining structural characteristics like edges and textures.
% With elaborate experiments, we show that the proposed downscaling method outperforms existing methods in preserving structure and minimizing artifacts. The proposed method can be extended for downscaling an image by a non-integer scaling factors;
%, overcoming limitations of traditional techniques and ensuring structural coherence for arbitrary scaling ratios, 
% which would be suitable for many real-world applications.
The main contributions of this
work are summarized as follows: 

\begin{enumerate}
    \item Our proposed methodology is, to the best of our knowledge, the most efficient kernel-based image downscaler, utilizing co-occurrence similarity across intensity pairs, while regulating the impact of co-occurrence similarity for each output pixel. The suggested approach incorporates all downscaling factors with minimal complexity appropriate for modern display devices. The principal characteristic of the suggested technique is its computational efficiency. 

\item  By executing this, we achieve results that are free from blurring and loss of intricate features. Experimental findings demonstrate a novel state-of-the-art classical downscaling technique that is applicable to various types of images, including nature scenes, cartoons, textures, and text.  To the best of our knowledge, this is the first method particularly designed and experimented for image downscaling by \textit{large scale}. 

\item We validate our approach on four standard datasets for varied downscaling to achieve state-of-the-art results. In particular, our proposed classical (non-deep) methodology offers competitive performance compared to a recent deep learning technique: SDFlow \cite{sun2023sdflow}. Our approach's computational efficiency makes it suitable for hardware and real-time applications like modern display devices.

% The results show that our method can improve the HR image reconstruction
% performance of the reduced LR image, compared with superresolution and encoder-decoder methods. In addition, compared
% with the reversible re-scaling network method, the proposed
% method can better alleviate the ill-posed problem for the im-
% age re-scaling task.

\end{enumerate}

\textbf{Organization:}
The structure of the remaining portion of the paper is as follows. In Section \ref{sec:related}, we provide a comprehensive review of the literature on image downscaling. A description of the proposed downscaling algorithm is provided in Section \ref{sec:method}. In Section \ref{sec:results}, we present the results of numerous investigations, which are highly competitive. We conducted exhaustive downscaling experiments with a diverse selection of image classes and downscaling factors to illustrate the efficacy of our proposal. Finally, we reach a conclusion in Section \ref{sec:conclusion}.

\section{RELATED WORK}
\label{sec:related}

A fundamental process in image processing, image downscaling is designed to reduce the spatial resolution of images while preserving perceptual quality. Applications like picture compression, transmission across networks with constrained bandwidth, and viewing on low-resolution screens all depend on it. The main difficulty is removing unnecessary information while maintaining crucial structural and textural details like edges and curves. From conventional interpolation-based techniques to contemporary deep learning models, a variety of current approaches are developing.

\subsection{Approaches Based on Interpolation}

Conventional downscaling techniques mostly use fixed-kernel interpolation approaches. One of the most popular methods is cubic interpolation~\cite{keys1981cubic} because of its ease of usage and comparatively decent visual performance. Compared to nearest-neighbor or bilinear interpolation, it produces smoother results by calculating the output pixel value using a weighted average of the surrounding pixels. However, it frequently fails to maintain edge sharpness and creates visible blurring in high-frequency regions. By utilizing a windowed sinc function as the kernel, Lanczos resampling~\cite{lanczos1964interpolation} enhances bicubic interpolation. Although this approach preserves image sharpness and frequency components better, it is less appropriate for real-time applications because to its higher computing complexity.

\subsection{Techniques Based on Optimization}

Downscaling is formulated as an inverse issue in optimization-driven techniques, which aim to minimize a loss function that gauges the perceptual similarity between the downscaled output and the high-resolution input. The application of the Structural Similarity Index (SSIM)~\cite{wang2004image} as an objective function to more accurately represent human visual perception is a well-known example.  Liu et al.~\cite{liu2018l0} presented a L0 gradient minimization technique that promotes image gradient sparsity. This method successfully maintains crucial image properties while downscaling by suppressing small variations while keeping critical edge structures. Authors in \cite{yang2024parameterized} presented a deep unsupervised learning-based solution to the $L_0$-regularized optimization issue. A single fully convolutional network is trained to solve the $L_0$ norm after it has been approximated using the $L_1$ norm with different parameters. A combined optimization technique that may concurrently learn arbitrary downscaling and arbitrary upscaling was proposed by the authors in \cite{pan2022towards}. 

\subsection{Models Based on Deep Learning}

The environment of image downscaling has changed dramatically as a result of recent developments in deep learning. Learning intricate mappings between high-resolution and low-resolution images is a remarkable skill that Convolutional Neural Networks (CNNs) have demonstrated. A study by Vasileiou et al.~\cite{vasileiou2022efficient} showed how CNN-based downscaling can improve resolution in millimeter-wave radar visuals.
Originally designed for super-resolution, Dong et al.'s SRCNN model~\cite{dong2016image, dong2016fast} has been modified for downscaling by flipping the training goal. By learning to map high-resolution patches to their low-resolution equivalents, the model produces an end-to-end solution that performs better in terms of perceptual quality than conventional techniques. 
Hybrid methods, like those put forth by Kim et al.~\cite{kim2018task}, blend deep learning refinement with conventional interpolation (such bicubic). By employing CNNs to improve perceptual qualities after downscaling using traditional methods, these approaches achieve a compromise between interpretability and performance.
To create pixels in the downscaled image, a resampler network, first presented in \cite{sun2020learned}, creates content adaptive image resampling kernels that are applied to the original HR input. 
% The authors of \cite{hou2018learning} downscale an input image to the desired size using a convolutional neural network (CNN) as a non-linear mapping function.  
A convolutional block architecture for fractional downsampling factor was presented by Chen et al. \cite{chen2022convolutional}. This design is crucial for many real-world image and video processing applications. Zhang et al. \cite{zhang2022enhancing} used a latent variable in their architecture to present a low-complexity invertible image downscaling model. In \cite{guo2023dinn360}, another invertible network that facilitates 360-degree picture downscaling was presented. Recently, \cite{li2025lightweight} presented a compression-aware picture rescaling technique.

Video downscaling techniques are being intensively researched by numerous researchers \cite{chen2024learned, xiang2022learning, cho2021learning}.
A progressive residual learning network architecture was introduced by Chen et al. \cite{chen2024learned} for video downsampling in streaming. The application of incorrect spatio-temporal downsampling to movies has been shown to result in aliasing problems \cite{xiang2022learning}. A neural network framework that simultaneously learns spatiotemporal downsampling and upsampling is proposed by the authors in \cite{xiang2022learning} to address this problem. A kernel-learning-based image downscaler that allows arbitrary downscaling factors through basic linear mapping was proposed by Cho et al. \cite{cho2021learning}.
In this field, encoder-decoder architectures are frequently employed, in which the encoder obtains a condensed representation of the image and the decoder reconstructs a downscaled version that is perceptually consistent. To increase realism, Generative Adversarial Networks (GANs) have also been used. Zhang et al.~\cite{zhang2018beyond}, for instance, presented a GAN-based system in which the generator learns to generate downscaled images that are visually convincing. Besides that, Authors in \cite{xia2020dsm} proposed a digital surface model (DSM)-based co-occurrence matrix (DSMB-CM) for semantic classification. Tian et al.~\cite{tian2025tree} proposed a tree-guided CNN for image super-resolution (TSRNet). Authors in \cite{wu2025dimension} proposed a sparse spectral spatial locality preserving projection method for hyperspectral image downscaling. Yao et al.~\cite{yao2022multi} proposed a MRSI matching method based on co-occurrence filter (CoF) space matching (CoFSM).

\begin{figure*}
    \centering
    \subfigure[Input.]{\includegraphics[width=0.19\textwidth]{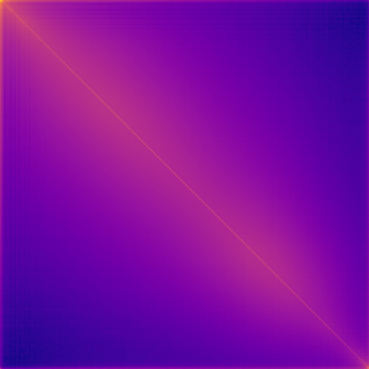}} 
    \subfigure[2x output.]{\includegraphics[width=0.19\textwidth]{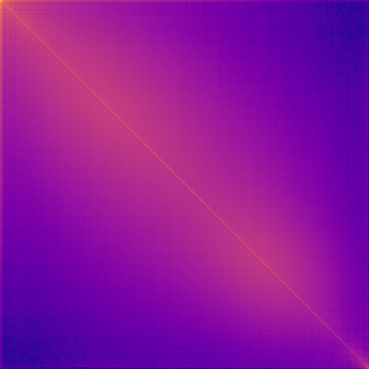}} 
    % \subfigure[IDCL,\textcolor{red}{(33.68, 98.9)}.]{\includegraphics[width=0.19\textwidth]{./images/IDCL_downscaled_peacockx2.png}}
     \subfigure[4x output.]{\includegraphics[width=0.19\textwidth]{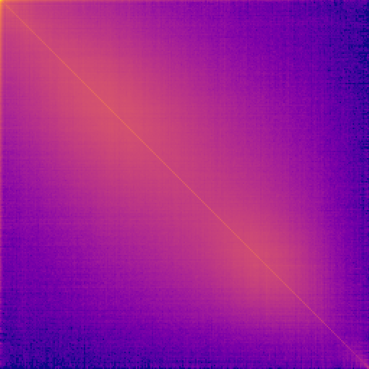}}
    \subfigure[8x output.]{\includegraphics[width=0.19\textwidth]{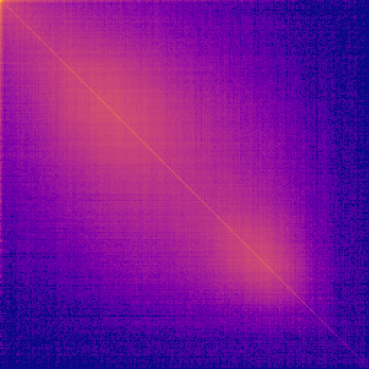}}
    \subfigure[16x output.]{\includegraphics[width=0.19\textwidth]{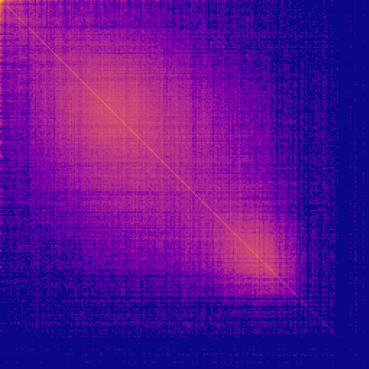}}
  \caption{Co-occurrence maps of input and downscaled images for differnt downscaling factors and for $k=3$ while computing the cco-occurrence maps It is important to note that the downscaled images exhibit similar pattern of co-occurrence profiles. In principle, we could also display the results for different value of $k$, for example $k=7$.}
    \label{fig:peacock_tonemap}
\end{figure*}

% Proposed Method
\section{Proposed Method}
\label{sec:method}
The proposed method, referred to as \textit{Large Scale Image Downscaling} (LSID), is designed to preserve important image details during resolution reduction by incorporating non-local co-occurrence statistics into a guided filtering framework. The algorithm operates in two main phases, co-occurrence profile learning, and Guided kernel filtering for downscaling.
% \begin{enumerate}
%     \item Co-occurrence profile learning, and
%     \item Guided kernel filtering for downscaling.
% \end{enumerate}
An intermediate guide image is generated via uniform averaging to provide a coarse approximation of the downscaled result, which is then refined using similarity weights derived from the learned co-occurrence matrix.

\subsection{Generation of Intermediate Image}
% We first compute an intermediate image $G$ of size $(H' \times W')$ using a $(2s+1) \times (2s+1)$ uniform averaging kernel applied over non-overlapping regions in $I$. This \textit{box filtering} step provides a crude approximation of the downscaled image, ensuring that local intensity trends are preserved. The intermediate downscale image namely guide image $G$ may contain non-integer pixel values; for subsequent co-occurrence processing, these values are rounded to the nearest integer.
Initially, we compute an intermediate representation $G$ with dimensions $(H' \times W')$ from input image with dimensions $(H \times W)$, where $H' = \frac{H}{s}$ and $W' = \frac{W}{s}$.
 This image is generated by employing a uniform averaging kernel of size $(2s+1) \times (2s+1)$ (where $s$ is a downscaling factor) on non-overlapping regions within $I$.  For a downscaling factor of 4, the dimensions of the averaging kernel will be $9 \times 9$. This \textit{box filtering} step offers a suboptimal approximation of the downscaled image, ensuring the preservation of local intensity trends. The intermediate downscale image, known to as the guide image $G$, may possess non-integer pixel values. To enhance subsequent co-occurrence processing, these numbers are rounded to the nearest integer.

\subsection{Co-occurrence Profile Learning}
The co-occurrence profile $C$ is a $256 \times 256$ matrix (for 8-bit images) that quantifies the frequency of occurrence of intensity pairs $(a,b)$ within a spatial neighborhood.
For each pixel $i$ in input image $I$, we identify its $K$-neighborhood $S(i)$, where $K$ is typically chosen as $K \geq s$. If $a$ denotes the intensity at pixel $i$, and $b$ denotes the intensity at a neighbor $j \in S(i)$, the co-occurrence matrix entry $C(a,b)$ is incremented. This process is repeated over all pixels, resulting in a statistical profile that captures non-local intensity relationships in the original image. The co-occurance map is computed as:
\begin{equation}
C(a, b) = \sum_{i,j} [I(i) = a] \, [I(j) = b]
\end{equation}
The notation $[\cdot]$ is defined such that it yields a value of 1 when the expression contained within the brackets is true, and a value of 0 when it is false. 
The neighborhood in (1) is characterized as 
\[
\mathcal{S}(i) = i + S, \quad \text{where} \quad S = [-K, K]^2.
\]
The computed $C$ is then normalized by its maximum value to obtain $C_{\text{norm}} \in [0,1]$, which will be used for adaptive kernel weighting in the next step. The normalized co-occurrence map can be computed as: 
\begin{equation}
C_{\text{norm}} = \frac{C(a,b)}{\max(C)}
\end{equation}
The co-occurrence maps of input image and downscaled images for scaling factors $2\times$, $4\times$, $8\times$ and $16\times$ is shown in Figure \ref{fig:peacock_tonemap}. Notice that the downscaled images have the co-occurrence profiles very similar to the input image. This is the intuitive yet fundamental basis of our proposed downscaling method. 

\subsection{Co-occurrence Guided Kernel Filtering}
The refined downscaling is achieved via a kernel filtering process where the weight of each contributing pixel is determined by the similarity between the guide pixel intensity $G(p)$ and the actual input pixel intensity $I(j)$, as encoded in $C_{\text{norm}}$.
For each output pixel $p$, we define an input mapping region $\Omega(p)$ corresponding to its spatial footprint in the input image. The output pixel value is computed as follows:
\begin{equation}
J(p) = \frac{\sum_{j \in \Omega(p)} w_{pj} \cdot I(j)}{\sum_{j \in \Omega(p)} w_{pj}}
\end{equation}
where
\begin{equation}
w_{pj} = \exp\left( \alpha \cdot C_{\text{norm}}(G(p), I(j)) \right)
\end{equation}
and $\alpha$ is a tunable parameter controlling the influence of co-occurrence similarity and and $C_{\text{norm}}(G(p), I(j))$ is the normalized co-occurrence frequency  of the intensity pair $(G(p), I(j))$ in the input image $I$. 
This formulation ensures that pixels in the input neighborhood that share strong co-occurrence relationships with the guide pixel contribute more heavily to the output, thereby retaining structural details such as edges and textures. The weight $w_{pj}$ has certain properties:

\begin{itemize}
    \item \textbf{Positivity :} The exponential map exhibits a strictly positive characteristic that leads to 
    \[
    w_{pj} > 0 \quad\text{for all } p,j.
  \]
    \item \textbf{Monotonicity in \(C_{\mathrm{norm}}\):} 
  For a fixed \(\alpha>0\), \(w_{pj}\) exhibits a strictly increasing behavior in relation to \(C_{\mathrm{norm}}\); conversely, for \(\alpha<0\), it demonstrates a strictly decreasing trend. In formal terms, \[ 
  \frac{\partial w_{pj}}{\partial C_{\mathrm{norm}}} = \alpha\, w_{pj}, 
  \] 
     where the sign of \(\alpha\) is significant.
    % \item \textbf{Scalability:} Works for integer and non-integer downscaling factors by appropriately setting $K$ and $\Omega(p)$.
    % \item \textbf{Channel Independence:} For RGB images, the process is applied independently to each color channel, enabling accurate preservation of chromatic details.
    \item \textbf{Log-linear representation and numerical convenience:}
  \[
    \log w_{pj} = \alpha\, C_{\mathrm{norm}}(G(p),I(j)).
  \]
  This representation proves to be advantageous when it is necessary to combine weights additively or when employing log-sum-exp for stable normalization.
\end{itemize}

% \subsection{Summary of Steps}
% The entire method is summarized below:
% \begin{enumerate}
%     \item \textbf{Input:} Image $I$ of size $(H \times W)$, scale factor $s$, co-occurrence neighborhood size $K$, parameter $\alpha$.
%     \item Generate guide image $G$ using uniform averaging with kernel size $(2s+1) \times (2s+1)$.
%     \item Compute normalized co-occurrence matrix $C_{\text{norm}}$ from $I$ using Algorithm~\ref{alg:cooc}.
%     \item Downscale using co-occurrence guided kernel filtering as described in Algorithm~\ref{alg:idcl}.
%     \item \textbf{Output:} Downscaled image $O$ of size $(H' \times W')$.
% \end{enumerate}

% \subsection{Advantages}
% The proposed IDCL framework offers the following benefits:
% \begin{itemize}
%     \item \textbf{Detail Preservation:} By incorporating co-occurrence similarity, the algorithm selectively enhances contributions from structurally similar pixels, mitigating blurring artifacts.
%     \item \textbf{Adaptivity:} The method adapts to local image statistics, unlike fixed interpolation schemes.
%     \item \textbf{Scalability:} Works for integer and non-integer downscaling factors by appropriately setting $K$ and $\Omega(p)$.
%     \item \textbf{Channel Independence:} For RGB images, the process is applied independently to each color channel, enabling accurate preservation of chromatic details.
% \end{itemize}

\begin{algorithm}[H]
\caption{Co-occurrence Matrix Computation}
\label{alg:cooc}
\begin{algorithmic}[1]
\State \textbf{Input:} Image $I$ of size $(H\times W)$; neighborhood size $K$
\State \textbf{Output:} Normalized co-occurrence matrix $C_{\text{norm}}$
\State $C \gets 0_{256\times256}$
\For{$i = 1,\dots,H\times W$}
  \State $a \gets I(i)$
  \For{$j \in S(i)$} \Comment{$S(i)$ denotes the $K$-neighborhood of pixel $i$}
    \State $b \gets I(j)$
    \State $C(a,b) \gets C(a,b) + 1$
  \EndFor
\EndFor
\State $C_{\text{norm}} \gets C / \max(C)$
\end{algorithmic}
\end{algorithm}

\begin{algorithm}[H]
\caption{Proposed LSID Algorithm}
\label{alg:idcl}
\begin{algorithmic}[1]
\State \textbf{Input:} Image $I$ of size $(H\times W)$; scale factor $s$; parameter $\alpha$; $C_{\text{norm}}$
\State \textbf{Output:} Downscaled image $O$ of size $(H'\times W')$
\State Generate intermediate image $G$ of size $(H'\times W')$ using average filtering with a $(2s+1)\times(2s+1)$ uniform kernel
\For{$p = 1,\dots,H'\times W'$}
  \State $P \gets 0$; $Q \gets 0$
  \For{$j \in \Omega(p)$} \Comment{$\Omega(p)$: input image region mapped to output pixel $p$}
    \State $w \gets \exp\big(\alpha\cdot C_{\text{norm}}(G(p),\,I(j))\big)$
    \State $P \gets P + w\cdot I(j)$
    \State $Q \gets Q + w$
  \EndFor
  \State $J(p) \gets P / Q$
\EndFor
\end{algorithmic}
\end{algorithm}

\begin{table*}%[h]
    \centering
    \caption{Quantitative evaluation results (PSNR / SSIM) for different downscaling factors for DIV2K dataset that contains ground truths for $2\times$, $4\times$, and $8\times$ downscaling factors. The \textbf{best} results are marked in bold.}
    \begin{tabular}{|c|p{2cm}|p{2cm}|p{2cm}|p{2cm}|p{2cm}|p{2cm}|}
        \hline
        {Scaling factor} & {Bicubic \cite{keys1981cubic}} & {Lanczos \cite{lanczos1964interpolation}} & {DPID \cite{weber2016rapid}} & {IDCL \cite{ghosh2023image}} & {SDFlow \cite{sun2023sdflow}} & {LSID (Ours)} \\
        \hline
        $2$ & 36.93 / 0.98 & 35.83 / 0.97 & 35.82 / 0.97 & 44.17 / 0.99 & -- & \textbf{44.34} / \textbf{0.99} \\
        $4$ & 30.47 / 0.94 & 26.72 / 0.84 & 32.98 / 0.95 & 36.37 / 0.98 & 33.62 / 0.94 & \textbf{39.22} / \textbf{0.98} \\
        $8$ & 23.92 / 0.87 & 23.98 / 0.87 & 26.62 / 0.91 & 29.97 / 0.96 & -- & \textbf{31.59} / \textbf{0.97} \\
        \hline
    \end{tabular}
    \label{tab:comparison}
\end{table*}

% updated table for d = 4

\begin{table*}%[h]
    \centering
    \caption{Quantitative results (PSNR / SSIM) for downscaling factor $d = 4$ on different datasets. The \textbf{best} results are marked in bold.}
    \begin{tabular}{|c|p{3cm}|p{3cm}|p{3cm}|}
        \hline
        {Downscaling Method} & DIV2K Dataset & BSD100 Dataset & Urban100 Dataset \\
        \hline
        Bicubic \cite{keys1981cubic} & 30.47 / 0.942 & 26.34 / 0.847 & 23.46 / 0.826 \\ 
        Lanczos \cite{lanczos1964interpolation} & 30.35 / 0.950 & 26.72 / 0.849 & 23.69 / 0.831 \\
        Content-adaptive \cite{kopf2013content} & 29.04 / 0.950 & 27.11 / 0.851 & 23.93 / 0.832 \\
        DPID \cite{weber2016rapid} & 32.98 / 0.950 & 39.57 / 0.989 & 36.20 / 0.986 \\
        IDCL \cite{ghosh2023image} & 36.37 / 0.980 & 36.27 / 0.979 & 32.12 / 0.971 \\
        ADL \cite{zhang2023adaptive} & 33.05 / 0.931 & 29.12 / 0.873 & 26.81 / 0.867 \\
        SDFlow \cite{sun2023sdflow} & 33.62 / 0.946 & 29.51 / 0.892 & 26.95 / 0.872 \\
        LSID (Ours) & \textbf{39.22} / \textbf{0.983} & \textbf{36.67} / \textbf{0.994} & \textbf{36.75} / \textbf{0.983} \\
        \hline
    \end{tabular}
    \label{tab:comparison_for_4}
\end{table*}

\begin{table*}%[h]
    \centering
    \caption{Quantitative no-reference quality assessment using NIQE, BRISQUE, and PIQE metrics (lower is better) for downscaling factor $16\times$. The \textbf{best} results are highlighted in bold.}
    \begin{tabular}{|c|c|c|c|}
        \hline
        \textbf{Downscaling Method} & \textbf{NIQE} $\downarrow$ & \textbf{BRISQUE} $\downarrow$ & \textbf{PIQE} $\downarrow$ \\
        \hline
        Bicubic & 18.87 & 40.53 & 29.74 \\ 
        IDCL & 18.88 & 39.63 & 39.64 \\ 
        DPID & 18.93 & 42.49 & 32.25 \\ 
        SDFlow & 18.90 & 45.67 & 35.12 \\ 
        LSID (Ours) & \textbf{18.84} & \textbf{38.13} & \textbf{26.35} \\ 
        \hline
    \end{tabular}
    \label{tab:niqe_brisque_piqe}
\end{table*}

\section{Experimental Results}
\label{sec:results}

\subsection{Dataset}

We tested a wide range of publicly accessible high-resolution image datasets, including DIV2K \cite{agustsson2017ntire}, BSD100 \cite{martin2001database}, Urban100 \cite{huang2015single}, and RealSR \cite{cai2019toward}. A well-known benchmark for image restoration and downscaling tasks is the DIV2K (Diverse 2K) dataset, which was first presented at the NTIRE 2017 Challenge on Single Image Super-Resolution \cite{agustsson2017ntire}.We performed experiments on several downscaling factors $d \in \{2, 4, 8, 16\}$. The DIV2K dataset \cite{agustsson2017ntire} comprises high-resolution images with complex high-frequency details, which complicate the retention of structural elements as the downscaling factor escalates, making it crucial to maintain details with the growing downscaling factor. The compilation of four datasets guarantees a comprehensive assessment across various image categories and complexity levels, as well as both fixed and arbitrary downscaling variables.

\subsection{Downscaling Performance Comparison}

We assessed our technique on various high-resolution image datasets to thoroughly evaluate its downscaling performance. The main aim was to assess the algorithm's capacity to maintain structure, texture, and perceived quality across a diverse range of images. For a specified scale factor, the technique was implemented on high-resolution input photos, and the resultant downscaled outputs were evaluated against either existing ground-truth references or conventional interpolation-based baselines. A quantitative performance evaluation was conducted using two image quality metrics: the Structural Similarity Index (SSIM) and the Peak Signal-to-Noise Ratio (PSNR) \cite{wang2004image}. We use non-reference image quality measurements because there is no ground truth available for downscaling factor $16\times$ in any of the datasets. Perception-based Image Quality Evaluator (PIQE) \cite{ganesan2020color}, Naturalness Image Quality Evaluator (NIQE) \cite{mittal2012making}, and Blind/Referenceless Image Spatial Quality Evaluator (BRISQUE) \cite{mittal2012no}.

We compare our proposed LSID method with recent classical techniques for image downscaling for quantitative evaluation. These techniques include bicubic interpolation \cite{keys1981cubic}, Lanczos resampling \cite{lanczos1964interpolation}, content-aware downscaling \cite{kopf2013content}, detail preserving image downscaling (DPID) \cite{weber2016rapid}, and image downscaling via co-occurrence learning (IDCL) \cite{ghosh2023image}. Additionally, we conduct a comparison with the most advanced deep learning techniques.{Adaptive downsampling models (ADL) \cite{son2022toward} and SDFlow \cite{sun2023sdflow} were used to downscale the image in deep learning methods.

In the DIV2K dataset \cite{agustsson2017ntire}, our LSID technique reaches a maximum PSNR of \textbf{39.22} dB and SSIM of \textbf{0.983}, while the recent SDFlow method (deep learning) yields a PSNR of 33.62 dB and SSIM of 0.946. The enhanced compilation of natural as well as high resolution images in the DIV2K dataset verifies that LSID excels in delivering cutting-edge downscaling of natural images. The intricate sceneries with repetitive patterns in the BSD100 \cite{martin2001database} and Urban100 \cite{huang2015single} datasets render the preservation of delicate geometric details in downscaled photos a formidable challenge. The comparatively diminished values of PSNR and SSIM for the BSD100 \cite{martin2001database} and Urban100 \cite{huang2015single} datasets are evident in Table~\ref{tab:comparison_for_4}. Additonaly, we attain BRISQUE and PIQE \textbf{38.13} and \textbf{26.35} while performing downscaling factor 16 on the DIV2K dataset.  Nevertheless, our LSID approach continues to demonstrate superior performance with respect to both PSNR and SSIM measures. The experimental results for downscaling factors $2\times$, $4\times$, $8\times$, and $16\times$ presented in Table~\ref{tab:comparison} and Table~\ref{tab:niqe_brisque_piqe} demonstrate the effectiveness of our LSID technique in maintaining both pixel-level and perceptual quality across the DIV2K high-resolution image dataset. We set $k=3$ and $\alpha=5$ for our LSID method to demonstrate our results.

\begin{figure*}
    \centering
    \subfigure[Input ($2040 \times 1584$).]{\includegraphics[width=0.23\textwidth]{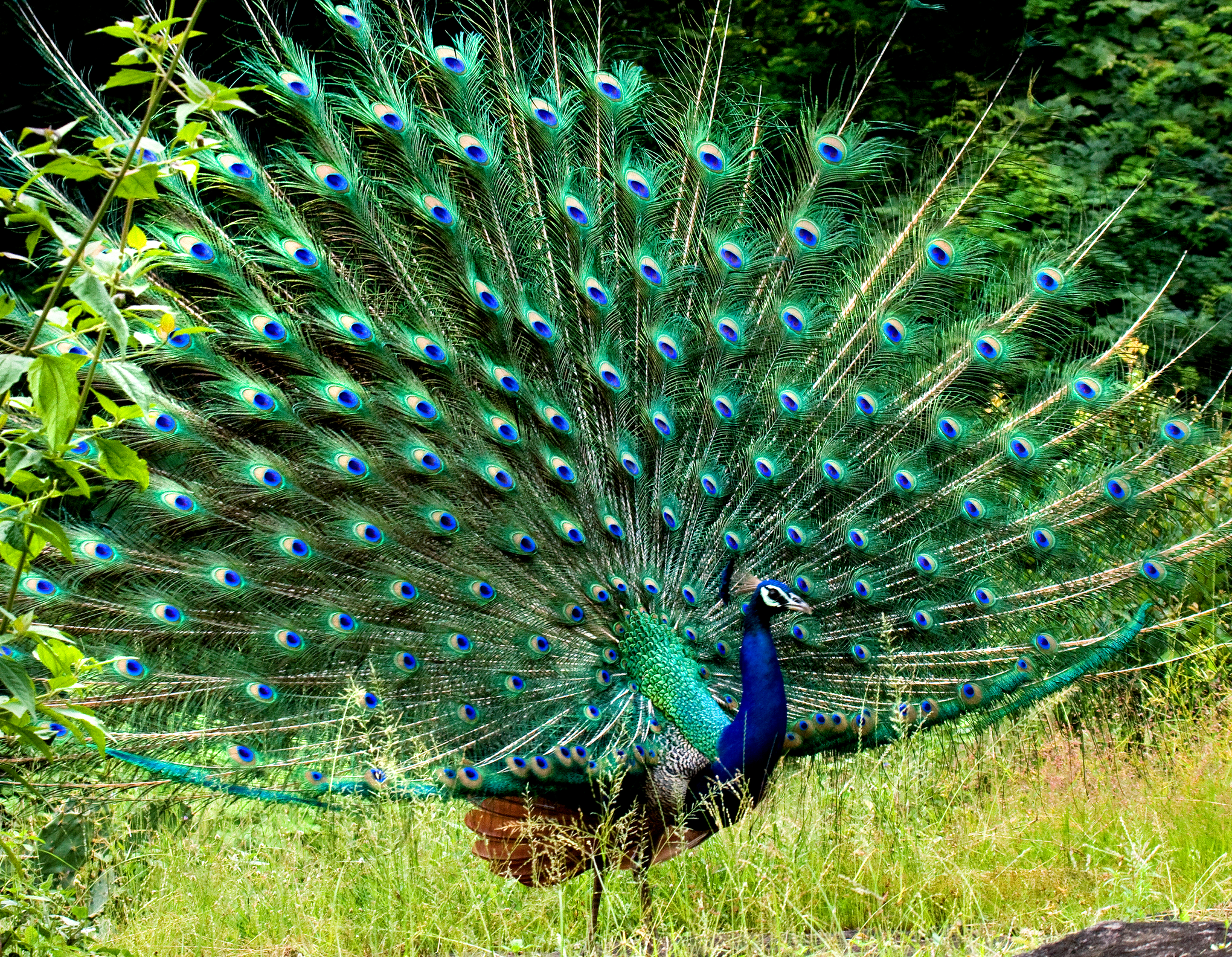}} 
    \subfigure[Bicubic, \textcolor{red}{(25.97, 94.3)}.]{\includegraphics[width=0.23\textwidth]{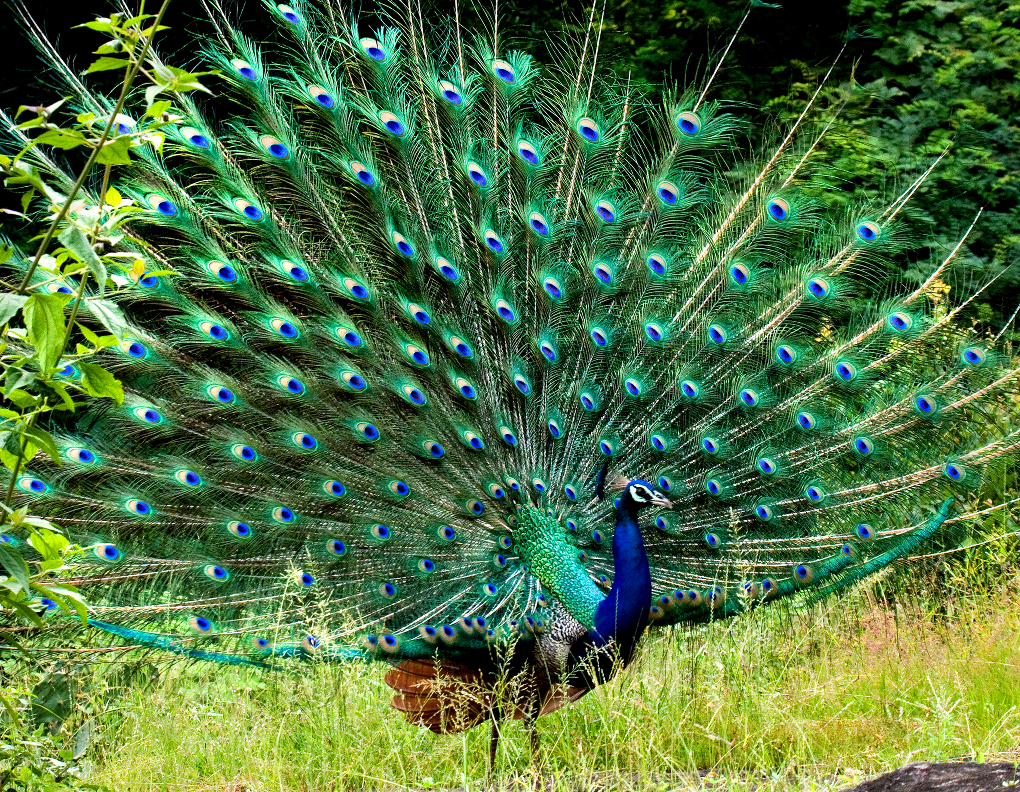}} 
    % \subfigure[IDCL,\textcolor{red}{(33.68, 98.9)}.]{\includegraphics[width=0.19\textwidth]{./images/IDCL_downscaled_peacockx2.png}}
     \subfigure[DPID, \textcolor{red}{(26.08, 94.9)}.]{\includegraphics[width=0.23\textwidth]{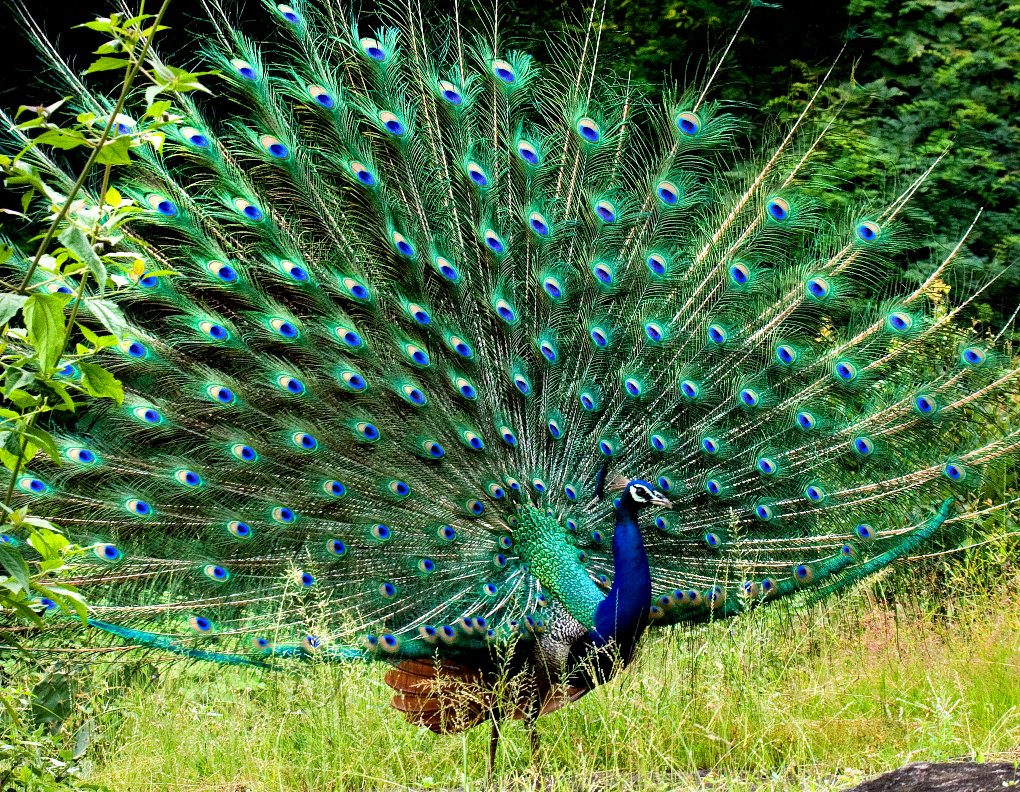}}
    \subfigure[LSID, \textcolor{red}{(33.70, 98.8)}.]{\includegraphics[width=0.23\textwidth]{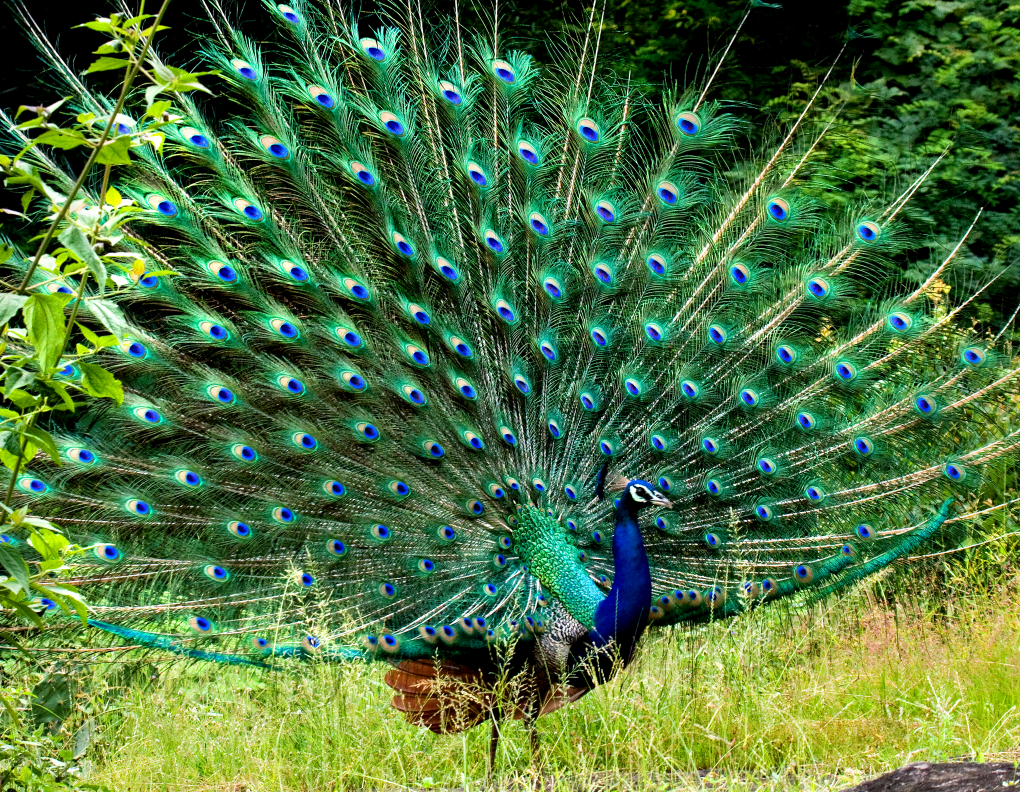}} \\
\texttt{Downscaling factor 2} \\~\\
   % \subfigure[Zoomed 16x.]{\includegraphics[width=0.18\textwidth]{./images/zoom_16x_crop.pdf}} 
   % \hspace{3.70cm}
    \subfigure[Bicubic, \textcolor{red}{(19.20, 78.2)}.]{\includegraphics[width=0.23\textwidth]{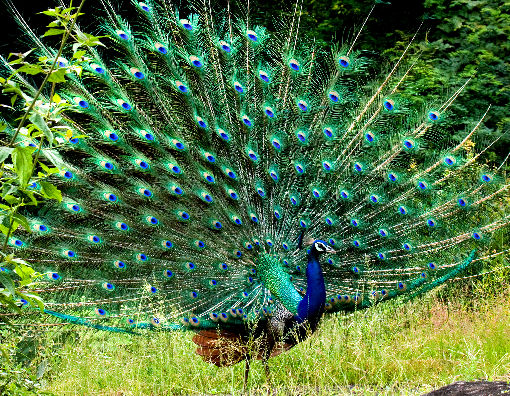}} 
    % \subfigure[IDCL,\textcolor{red}{(33.15, 98.5)}.]{\includegraphics[width=0.19\textwidth]{./images/IDCL_downscaled_x4_peacock.png}}
    \subfigure[SDflow, \textcolor{red}{(18.99, 67.8)}]{\includegraphics[width=0.23\textwidth]{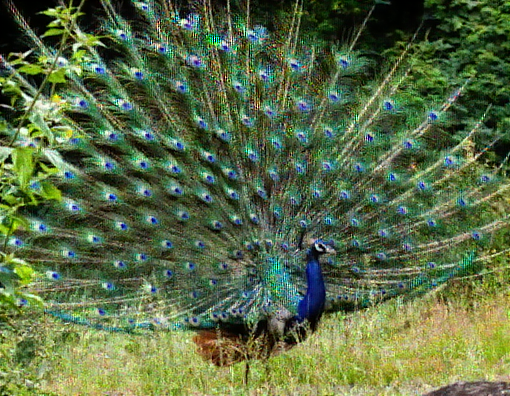}}
    \subfigure[LSID, \textcolor{red}{(33.17, 98.5)}.]{\includegraphics[width=0.23\textwidth]{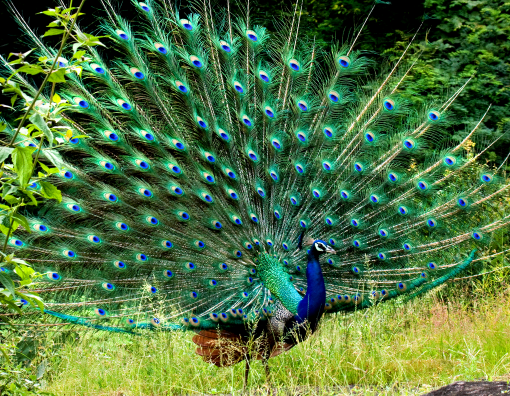}} \\
    \texttt{Downscaling factor 4} \\~\\
  %  \hspace{3.70cm}
    \subfigure[Bicubic, \textcolor{red}{(16.66, 64.8)}.]{\includegraphics[width=0.23\textwidth]{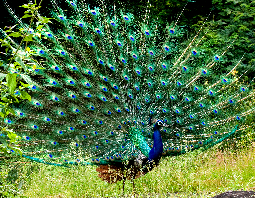}} 
   % \subfigure[IDCL,\textcolor{red}{(33.55, 98.5)}.]{\includegraphics[width=0.19\textwidth]{./images/IDCL_downscaled_x8_peacock.png}}
    \subfigure[DPID, \textcolor{red}{(23.60, 89.8)}.]{\includegraphics[width=0.23\textwidth]{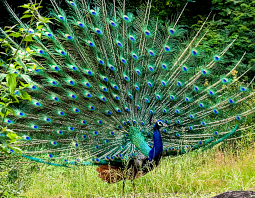}} 
    \subfigure[LSID, \textcolor{red}{(33.58, 98.5)}.]{\includegraphics[width=0.23\textwidth]{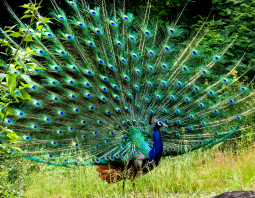}} \\
     \texttt{Downscaling factor 8} \\~\\
    % \hspace{3.70cm}
    \subfigure[Bicubic.]{\includegraphics[width=0.23\textwidth]{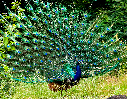}} 
    % \subfigure[IDCL.]{\includegraphics[width=0.19\textwidth]{./images/IDCL_downscaled_x16_peacock.png}}
    \subfigure[SDflow.]{\includegraphics[width=0.23\textwidth]{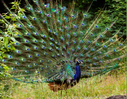}}
    \subfigure[LSID.]{\includegraphics[width=0.23\textwidth]{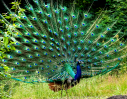}} \\
    \texttt{Downscaling factor 16} \\~\\
    \subfigure[Input.]{\includegraphics[width=0.107\textwidth]{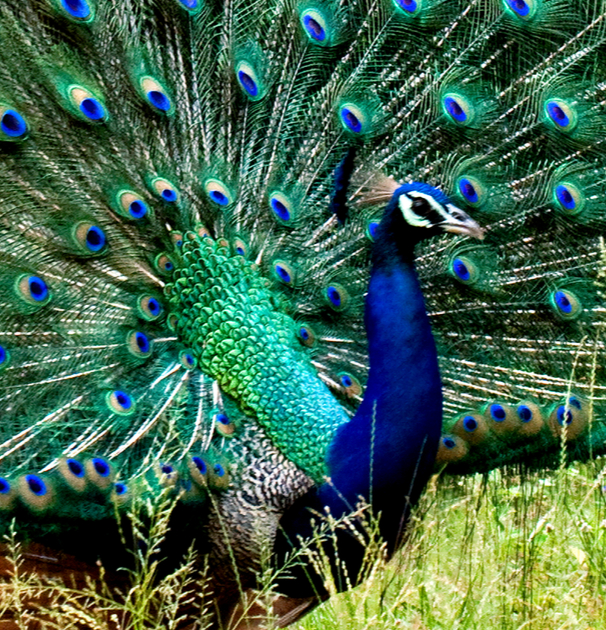}} 
\subfigure[Bicubic.]{\includegraphics[width=0.11\textwidth]{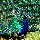}}  
\subfigure[SDflow.]{\includegraphics[width=0.11\textwidth]{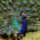}}
\subfigure[LSID.]{\includegraphics[width=0.11\textwidth]{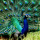}}  
\\
    \texttt{Zoomed view of the images for downscaling factor $16$.}
\\~\\
    \caption{Comparison of visuals across various downscaling factors analysis of $2\times$, $4\times$, $8\times$, and $16\times$ on the \textit{Peacock Image} from the DIV2K dataset, along with a comparison of various downscaling methods: Bicubic \cite{keys1981cubic}, DPID \cite{weber2016rapid}, SDflow \cite{sun2023sdflow} and our method LSID. The approach we utilize LSID effectively preserves structures such as peacock's feathers as the downscaling factor increases.}
    \label{fig:peacock_downscale_comp}
\end{figure*}

\subsection{Visual Comparisons Across Various Image Types}

In this section, we conduct a comprehensive qualitative analysis by assessing our method across four distinct image categories: natural sceneries, texture-dense images, text-rich content, and cartoon illustrations. This enables us to evaluate the extent to which LSID generalizes across a variety of visually diverse contexts and use cases.

\begin{figure*}
    \centering
    \subfigure[Input ($2040 \times 1356$).]{\includegraphics[width=0.19\textwidth]{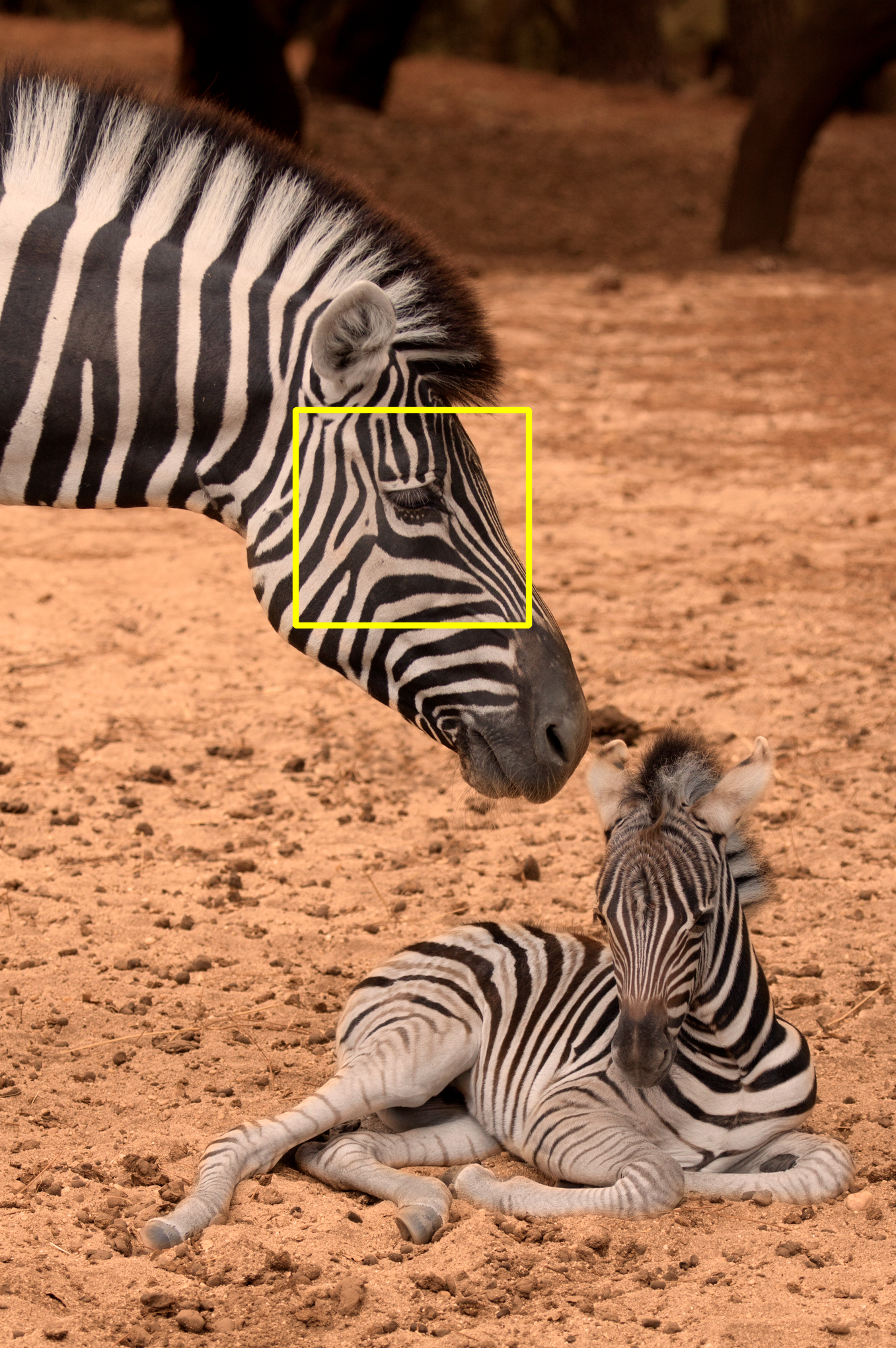}} 
    \subfigure[Bicubic, \textcolor{red}{(23.92, 87.5)}.]{\includegraphics[width=0.19\textwidth]{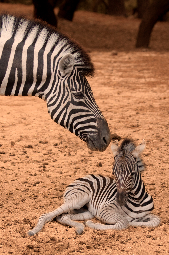}} 
    \subfigure[IDCL, \textcolor{red}{(29.97, 96.6)}.]{\includegraphics[width=0.19\textwidth]{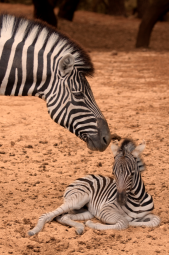}}
    \subfigure[SDflow, \textcolor{red}{(23.87, 87.8)}]{\includegraphics[width=0.19\textwidth]{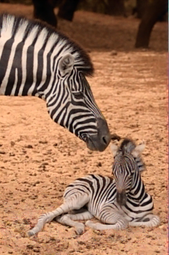}}
    \subfigure[LSID, \textcolor{red}{(31.59, 97.5)}]{\includegraphics[width=0.19\textwidth]{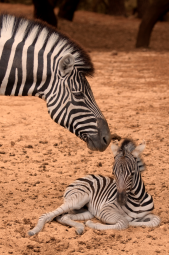}} \\
    \subfigure[Zoomed Input.]{\includegraphics[width=0.15\textwidth]{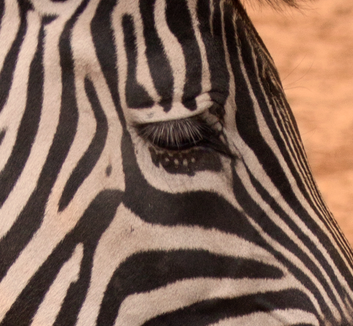}} 
    \subfigure[Zoomed Bicubic.]{\includegraphics[width=0.15\textwidth]{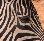}} 
    \subfigure[Zoomed IDCL.]{\includegraphics[width=0.15\textwidth]{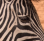}}
    \subfigure[Zoomed SDflow.]{\includegraphics[width=0.15\textwidth]{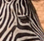}}
    \subfigure[Zoomed LSID.]{\includegraphics[width=0.15\textwidth]{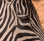}}
  %  \caption{(a) blah (b) blah (c) blah (d) blah}
  \caption{Visual comparison for the downscaling factor of $8$ on \textit{Zebra} image. We compare with different image downscaling techniques: Bicubic \cite{keys1981cubic}, IDCL \cite{ghosh2023image}, SDflow \cite{sun2023sdflow} and our method LSID for the downscaling factor of $8$. Our method LSID efficiently retains the sharpness and realistic pattern. We observe that we achieved the PSNR value = $31.59$ and SSIM = $97.5$.}
    \label{fig:zebra_comparison}
\end{figure*}

\subsubsection{\textbf{High Frequency Content}}

High-frequency images are distinguished by the rapid variations in pixel intensity that occur across small spatial regions. This means that the brightness or color values of individual pixels vary substantially. These swift intensity changes correspond to visually distinct characteristics such as edges, fine textures, patterns, and detailed details present in items like feathers, grass, or fabric. In Figure \ref{fig:peacock_downscale_comp}, we demonstrate the visual outcome of dowscaling a natural image, denoted \textit{peacock}, using our LSID method and other recent methods for downscaling factors $2\times$, $4\times$, $8\times$ and $16\times$ and we discovered that the majority of the compared methods tend to obscure essential details when high-frequency components, such as edges and textures, are present. With the increase of the downscaling factor and our methodology, DPID, Bicubic, and SDflow (which is dataset-specific) introduce more distortion in the peacock feather structures and details. LSID demonstrates fewer ringing or filtering effects and retains a greater amount of the original structure.

The preservation of fine edges and structural details becomes increasingly challenging as the downscaling factor increases, due to the significant loss of spatial information. In such high-reduction scenarios, conventional interpolation-based methods such as bicubic often produce overly smoothed results, while learning-based approaches like sdflow may fail to consistently maintain high-frequency textures. In contrast, the proposed LSID framework demonstrates robust performance even under large downscaling ratios. As illustrated the last row in Figure. \ref{fig:peacock_downscale_comp}, for a downscaling factor of 16, LSID effectively preserves intricate feather patterns and edge continuity in the peacock image, whereas noticeable blurring and structural degradation are evident in the outputs of bicubic and sdflow. This highlights the superior capability of LSID in maintaining visual fidelity and structural integrity under extreme compression conditions.

\begin{figure*}
    \centering
    \subfigure[Input ($480 \times 320$).]{\includegraphics[width=0.24\textwidth]{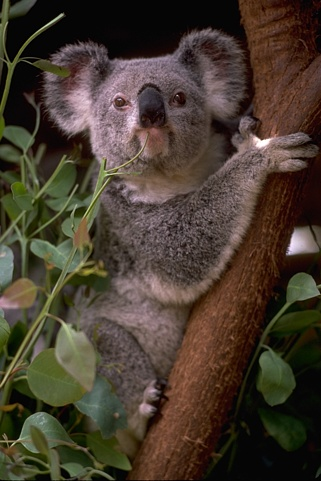}} 
    \subfigure[Bicubic, \textcolor{red}{(28.99, 91.30)}.]{\includegraphics[width=0.24\textwidth]{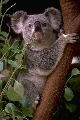}} 
    \subfigure[SDflow, \textcolor{red}{(25.02, 82.8)}]{\includegraphics[width=0.24\textwidth]{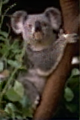}}
    \subfigure[LSID, \textcolor{red}{(35.77, 97.9)}]{\includegraphics[width=0.24\textwidth]{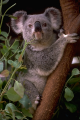}}
  \caption{Visual comparison for the downscaling factor of $4$ on \textit{Panda} image. We compare with different image downscaling techniques: bicubic \cite{keys1981cubic}, content-daptive \cite{kopf2013content}, DPID \cite{weber2016rapid},  SDflow \cite{sun2023sdflow} and our method LSID for the downscaling factor of $4$. Our method LSID efficiently retains the sharpness and realistic pattern. We observe that we achieved the PSNR value = $35.77$ and SSIM = $97.9$.}
    \label{fig:panda_comparison}
\end{figure*}

\subsubsection{\textbf{Downscaling Result on Natural Image}}

Complex structures, intricate patterns, and refined textures—such as animal fur, tree foliage, or skin tones—are often seen in natural photos, all of which enhance the perceived reality of the work. Downscaling these images is particularly challenging as it requires the retention of a wide range of spatial frequencies without the incorporation of artificial enhancements or dithering. Figure \ref{fig:zebra_comparison} illustrates the downscaling results for the \textit{zebra image} with a downscaling factor of $8\times$ and we also observe distortion, undesired halo effects, and edge ringing, particularly near high-contrast boundary lines. Notice that our proposed method (LSID) outperforms IDCL in terms of both PSNR and SSIM values. The superiority of our method is validated by the highest values of (PSNR, SSIM). LSID is involved in the preservation of structural content and the highest possible degree of similarity to the ground-truth image. 

% In Figure \ref{fig:panda_comparison}, we illustrate the downscaling results for downscaling factor of $4\times$ on \textit{Panda image} from BSD100 dataset and we observe that our method LSID preserves more structural details and less blurring effects on edges than SDflow and bicubic.

Figure \ref{fig:panda_comparison} presents the results of the $4\times$ downscaling applied to the \textit{Panda image} sourced from BSD100 dataset. LSID demonstrates a superior ability to maintain intricate structural details, such as the individual strands of fur, the precise contours of the ear and nose, and the textures of the leaves in the foreground, when compared to both bicubic and SDFlow methods. Bicubic results in a clear smoothing effect on small details and softens edges, leading to a reduction in high-frequency content. In contrast, SDFlow appears blurrier and shows more irregular artifacts in textured areas.

\begin{figure*}
    \centering
    \subfigure[Input ($2040 \times 1728$).]{\includegraphics[width=0.24\textwidth]{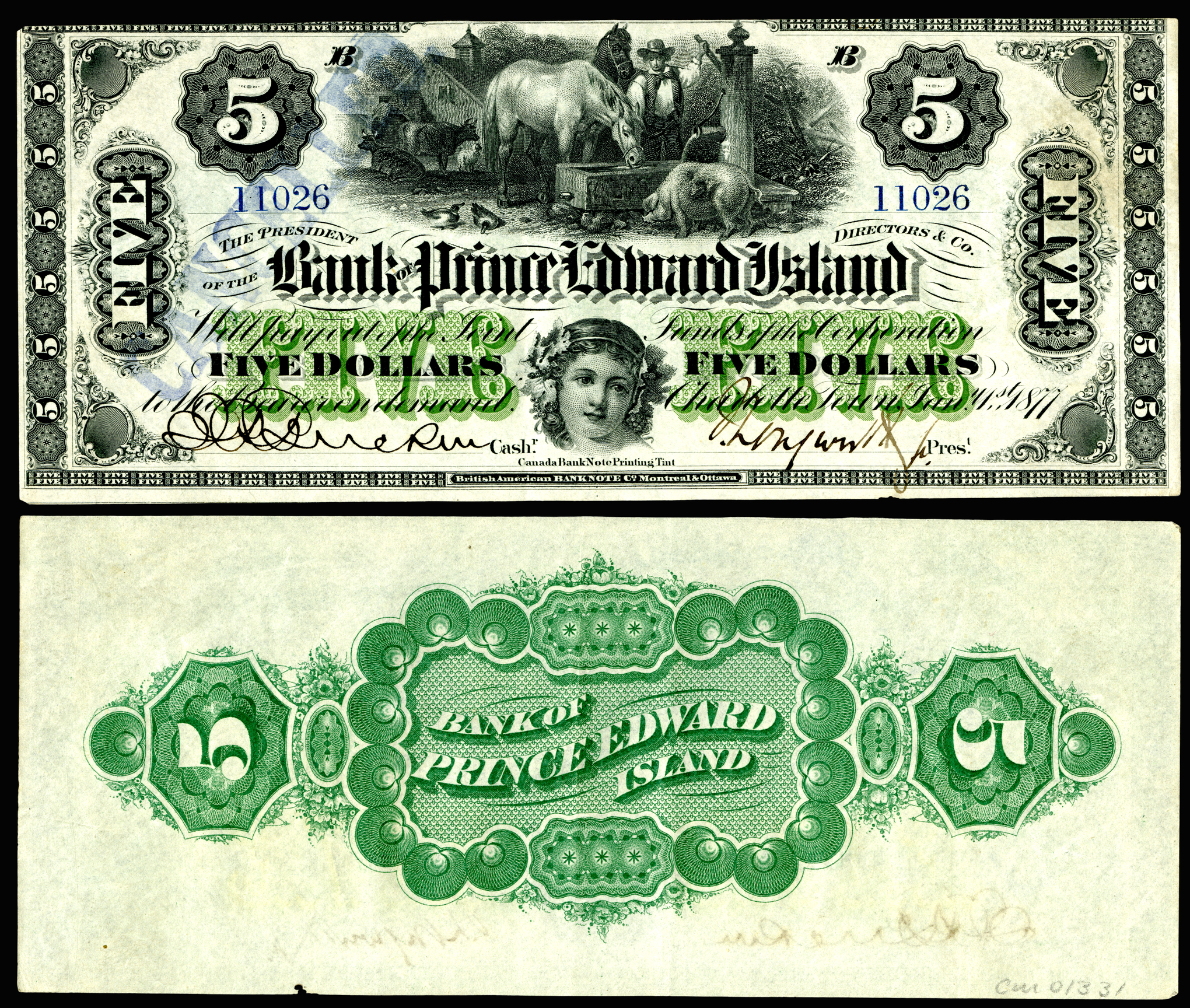}} 
    \subfigure[Bicubic.]{\includegraphics[width=0.24\textwidth]{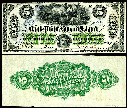}} 
  % \subfigure[DPID.]{\includegraphics[width=0.18\textwidth]{./images/0700x16_DPID_dollar.png}} 
  % \subfigure[Lanczos.]{\includegraphics[width=0.24\textwidth]{./images/0700x16_lanczos_dollar.png}}
    \subfigure[SDflow.]{\includegraphics[width=0.24\textwidth]{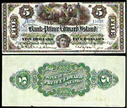}}
    \subfigure[LSID.]{\includegraphics[width=0.24\textwidth]{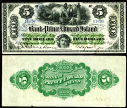}} 
\\
    \subfigure[Zoomed Input.]{\includegraphics[width=0.18\textwidth]{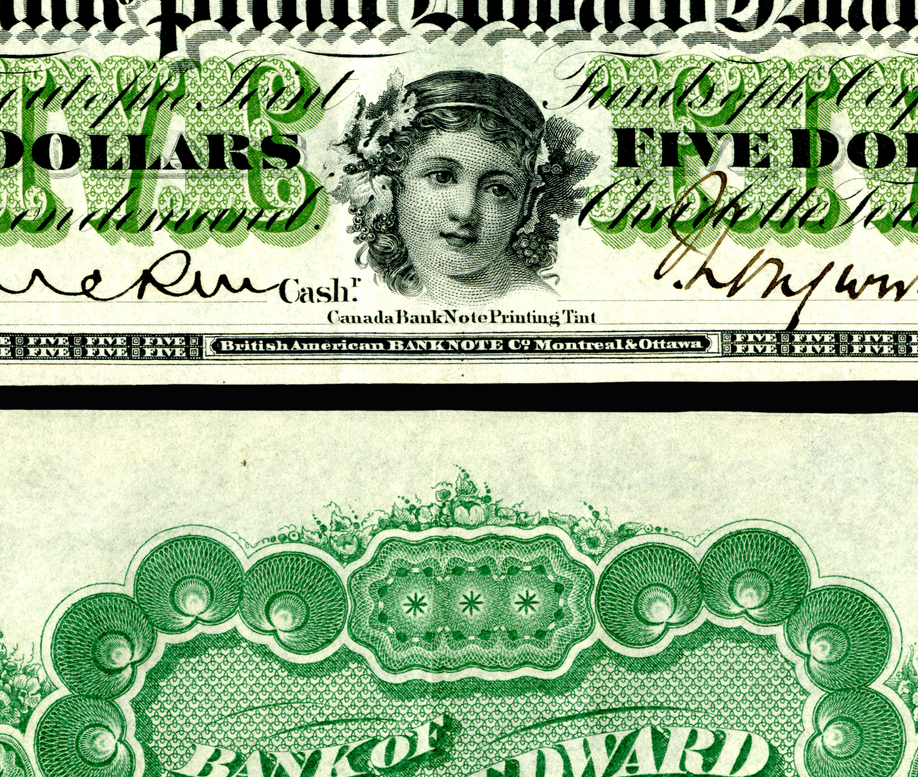}} 
    \subfigure[Zoomed Bicubic.]{\includegraphics[width=0.18\textwidth]{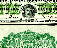}} 
  % \subfigure[Zoomed DPID.]{\includegraphics[width=0.18\textwidth]{./images/zoomed_DPID16_dollar.png}} 
  % \subfigure[Lanczos.]{\includegraphics[width=0.24\textwidth]{./images/0700x16_lanczos_dollar.png}}
    \subfigure[Zoomed SDflow.]{\includegraphics[width=0.18\textwidth]{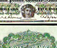}}
    \subfigure[Zoomed LSID.]{\includegraphics[width=0.18\textwidth]{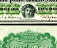}}
  %  \caption{(a) blah (b) blah (c) blah (d) blah}
  \caption{Visual comparison with different image downscaling techniques: bicubic \cite{keys1981cubic}, content-daptive \cite{kopf2013content}, DPID \cite{weber2016rapid}, IDCL \cite{ghosh2023image}, SDflow \cite{sun2023sdflow} our method LSID. For the downscaling factor of $16$, our method LSID achieves state-of-the-art performance in terms of perceptual metrics and visual detail retention. Furthermore, we observe that our method LSID delivers competitive perceptual quality.}
    \label{fig:dollar_down_16_comparison}
\end{figure*}

\subsubsection{\textbf{Downscaling Result on Text Image}}

When dealing with intricate ornamental patterns, fine textual details, and subtle tonal variations, such as those found in currency notes, image downscaling is particularly challenging. This is due to the fact that these elements must preserve both structural and stochastic components without the introduction of artificial smoothing or halo artifacts. In Figure \ref{fig:dollar_down_16_comparison}, we present the downscaling results for the downscaling factor of $16\times$ using our LSID and other methods on \textit{dollar image}. We observe that our LSID method preserves more structural details than the other methods, which are illusory by zooming results.

\subsection{Non-integer Downscaling of Images}

Most of the existing downscaling methods focussed on a relatively low to moderate scaling factor for image downscaling. However, in real time applications, it is often the case that we need to resize down an image by large scale, including a non-integer value. This is what we have particularly addressed in this work. We note that our proposed downscaling method is able to also perform resizing down an image by a non-integer factor. This is executed by first selecting the neighborhood using \textit{ceiling} value of the non-integer factor and then adapting the scheme of co-occurrence guided kernel filtering.

The effectiveness of the proposed method is demonstrated through a downscaling operation utilizing non-integer downscaling factors, which are relevant in practical applications. The downscaling was performed using the methodology for non-integer scale factors of 8.75 and 12.5. The results in Figure \ref{fig:sunflower} show that the downscaled images exhibit visually appealing characteristics, including smooth textures and well-preserved structural details. The method successfully maintains edge sharpness and overall visual fidelity, even with fractional scaling ratios, highlighting its robustness and adaptability for arbitrary non-integer downscaling scenarios.

\begin{figure*}
    \centering
    \subfigure[Input ($2040 \times 1152$).]{\includegraphics[width=0.32\textwidth]{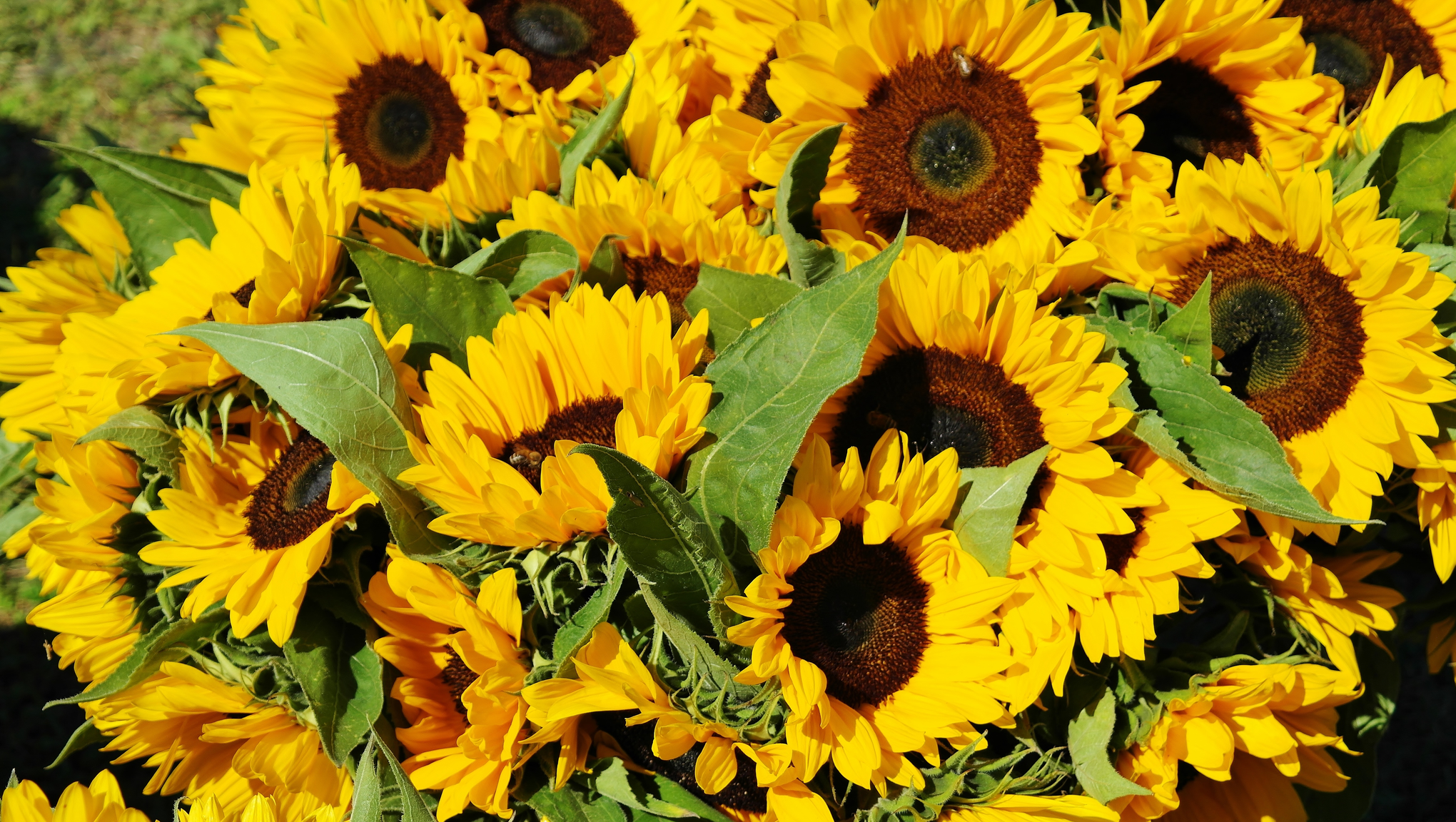}} 
   %  \subfigure[d = 2.7, \textcolor{red}{(NIQE = 3.80)}.]{\includegraphics[width=0.32\textwidth]{./images/LSID_X2.7.png}} 
   % \subfigure[d = 3.5, \textcolor{red}{(NIQE = 3.46)}.]{\includegraphics[width=0.32\textwidth]{./images/LSID_X3.5.png}}
   \subfigure[Downscaling factor $d = 8.75$.]{\includegraphics[width=0.32\textwidth]{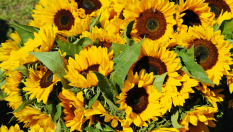}} 
   \subfigure[Downscaling factor $d = 9.50$.]{\includegraphics[width=0.32\textwidth]{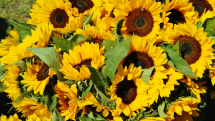}}
  %  \caption{(a) blah (b) blah (c) blah (d) blah}
  \caption{Visual results for non-integer downscaling factors: $8.75$ and $9.5$ and we observed that the NIQE values are 7.25 and 8.62 respectively. Notice that both downscaled images appear to be aesthetically pleasing with smooth textures and well-preserved structural details.}
    \label{fig:sunflower}
\end{figure*}

\section{Conclusion}
In this paper, we presented a new image downscaling technique LSID that utilizes co-occurrence similarity to inform the resizing process. Through the calculation of structural co-occurrence among pixels, LSID strategically enhances the contributions from locally similar neighborhoods while diminishing the impact of inconsistent or noisy neighbors. This approach effectively mitigates the prevalent issues of blurring and texture-washout that are often caused by fixed interpolation kernels. The algorithm functions autonomously on each RGB channel, allowing it to maintain intricate chromatic details and prevent cross-channel bleeding that could lead to de-saturation or blurring of color edges.
We demonstrated the effectiveness of our method for downscaling of images by large as well as non-integer downscaling factors. 

In future work, we plan to extend our method for video processing applications. This extension will facilitate the display of large images and videos on smaller devices. The high-quality downscaled images will effectively facilitate faster broadcasting and require reduced storage capacity.
\label{sec:conclusion}

\bibliographystyle{IEEEbib}
% \clearpage
\bibliography{strings}
\end{document}